\documentclass[11pt]{article}
\usepackage{amssymb}
\usepackage{citesort}

\renewcommand{\theequation}{\arabic{section}.\arabic{equation}}

\begin{document}
\begin{titlepage}
\thispagestyle{empty}

\begin{flushright}
                arXiv:0711.1619
\end{flushright}

\vskip 1cm

\begin{center}

{\LARGE\bf\sf Elliptic recurrence representation
\\[8pt]
of the $N=1$ Neveu-Schwarz blocks} \\

\end{center}

\vskip 1cm

\begin{center}

    {\large\bf\sf
    Leszek Hadasz${}^\dag$\footnote{\emph{e-mail}: hadasz@th.if.uj.edu.pl}$\!\!\!\!,\ \,$
    Zbigniew Jask\'{o}lski${}^\ddag$\footnote{\emph{e-mail}: jask@ift.uni.wroc.pl}
    and
    Paulina Suchanek${}^\dag$\footnote{\emph{e-mail}: suchanek@th.if.uj.edu.pl}
    }
     \\
\vskip 10mm
    ${}^\dag$ M. Smoluchowski Institute of Physics,
    Jagiellonian University, \\
    Reymonta 4,
    30-059~Krak\'ow, Poland, \\

\vskip 3mm
    ${}^\ddag$ Institute of Theoretical Physics,
    University of Wroc{\l}aw, \\
    pl. M. Borna, 50-204~Wroc{\l}aw, Poland. \\
\end{center}

\vskip 2cm

\begin{abstract}
We apply a suitably generalized method of Al.~Zamolodchikov to derive an elliptic
recurrence representation of the Neveu-Schwarz superconformal blocks
\end{abstract}

\vspace{\fill}


\vspace*{5mm}
\end{titlepage}

\section{Introduction}

Conformal field theory proved to be very efficient tool in
describing second order phase transitions in two-dimensional system
and is a commonly used language of string theory. Correlation
functions in CFT can be expressed as sums (or integrals) of
three-point coupling constants and the conformal blocks, fully
determined by the symmetry alone  \cite{Belavin:1984vu}.
  An explicit calculation of conformal blocks
is still one of the most difficult problems in CFT.
Not only the form of a general conformal block is unknown, but also its analytic
properties are still  conjectures rather than theorems.

On the other hand,
there exist very efficient recursive methods of an approximate, analytic determination
of a general 4-point conformal block
\cite{Zamolodchikov:ie,Zamolodchikov:2,Zamolodchikov:3}.
They were used for instance in checking the conformal bootstrap in the Liouville theory with the DOZZ
coupling constants \cite{Zamolodchikov:1995aa}, in study of the $c\to 1$ limit of minimal models
\cite{Runkel:2001ng} or in obtaining new results in the classical geometry of hyperbolic surfaces
\cite{Hadasz:2005gk}. In a more general context of an arbitrary CFT model these methods
allow for efficient  numerical calculations of any 4-point function once the structure constants are known.

In the last year the recursion representations have been worked out
for the super-conformal blocks related to the Neveu-Schwarz algebra
\cite{Hadasz:2006sb,Belavin:2006,Belavin:2007,Belavin:2007eq}. The so called
elliptic recursion was conjectured in \cite{Belavin:2007} for one type of NS blocks
and applied in the numerical
verification of the consistency of $N =1$ super-Liouville theory.
The extension of this method to another type of NS blocks was proposed
in \cite{Belavin:2007eq} where also further numerical support for the consistency of the
$N =1$ super-Liouville theory was given.

The aim of the present paper is to provide a comprehensive derivation
of the elliptic recursion for all types of NS blocks. This is done by
an appropriate extension of the method originally developed in
\cite{Zamolodchikov:2,Zamolodchikov:3} for the Virasoro case.
In our derivation we also use an exact analytic expressions
for certain $N=1$ NS superconformal blocks of $ c =\frac32$
theory obtained in \cite{hjs}.

The organization of the paper is as follows. In Section 2 we present our
notation and basic properties of NS blocks derived in \cite{Hadasz:2006sb}.
Section 3 is devoted to the analysis of the classical limit of
$N=1$ NS supersymmetric Liouville theory. Using the path integral representation
we show that  in the classical limit of the supersymmetric
Liouville correlators the leading terms are described by the classical bosonic Liouville action. This
implies in particular that the exponential part of the classical limit
of all NS blocks is given by the classical conformal block of the Virasoro theory.

One of the main points of the method proposed in \cite{Zamolodchikov:2,Zamolodchikov:3}
is that the dependence of the first two  terms of the ${1\over \Delta}$ expansion of quantum conformal block
on the external weights and the central charge
can be read off from the ${1\over \delta}$ expansion  of the classical block.
Since the extension of this reasoning to the NS N=1 case is rather straightforward
we present it in Section 4 mainly for completeness.

At this point one could use the results of
\cite{Zamolodchikov:2,Zamolodchikov:3} to derive the large $\Delta$ asymptotics
of  NS blocks. On the other hand one can follow the general strategy
of \cite{Zamolodchikov:2,Zamolodchikov:3} within the NS theory. This is done
in Section 5. The fact that the null vector decoupling equations of NS theory imply
exactly the same equation for the classical conformal block that one gets in the Virasoro case
can be seen as a consistency check of the path integral arguments
used in Section~3.

Finally in Section 6, using the explicit analytic expressions for  $c=\frac32$
superconformal blocks \cite{hjs} we derive the recurrence relations
for all type of NS superconformal blocks.

Some technical details of the elliptic Ansatz
used in \cite{Zamolodchikov:2,Zamolodchikov:3} are given in Appendix A.
In Appendix B it is shown that the recurrence formulae are in
perfect agreement with an exact analytical form
of the conformal blocks obtained in \cite{hjs}.

\section{NS $N=1$ superconformal blocks}

The  4-point NS superconformal blocks are conveniently defined in
terms of the 3-point block
$$
\rho{^{\Delta_3}_\infty}{^{\Delta_2}_{\:x}}{^{\Delta_1}_{\;0}}
: {\cal V}_{\Delta_3}\times {\cal V}_{\Delta_2} \times {\cal V}_{\Delta_1} \ \mapsto \ \mathbb{C},
$$
normalized by the condition
\[
\rho^{\Delta_3\ \Delta_2 \ \Delta_1}_{\infty \ \ x \ \ \ 0}(\nu_3,\nu_2,\nu_1)
\; = \;
\rho^{\Delta_3\ \Delta_2 \ \Delta_1}_{\infty \ \ x \ \ \ 0}(\nu_3,*\nu_2,\nu_1) \; = \; 1,
\]
where $\nu_3,\nu_2,\nu_1$ are super-primary states in NS superconformal Verma
modules
${\cal V}_{\Delta_3}, {\cal V}_{\Delta_2}, {\cal V}_{\Delta_1}$
and $*\nu_2=S_{-{1\over 2}}\nu_2$ \footnote{We shall follow the notation of \cite{Hadasz:2006sb}}.
Then the even,
\begin{eqnarray*}
\mathcal{F}^1_{\Delta}\!
\left[^{\underline{\hspace{3pt}}\,\Delta_3 \;\underline{\hspace{3pt}}\,\Delta_2}_{\hspace{3pt}\,\Delta_4 \;\hspace{3pt}\, \Delta_1} \right]\!(x)
    &=&
x^{\Delta - \underline{\hspace{3pt}}\,\Delta_2 - \Delta_1}
\left(
1 + \sum_{m\in \mathbb{N}} x^m\, F^m_{c, \Delta}\!
    \left[^{\underline{\hspace{3pt}}\,\Delta_3\;\underline{\hspace{3pt}}\,\Delta_2}_{\hspace{3pt}\,\Delta_4\;\hspace{3pt}\, \Delta_1} \right]
 \right),
\end{eqnarray*}
and the odd,
\begin{eqnarray}
\label{odd_block_def}
\mathcal{F}^{\frac{1}{2}}_{\Delta}\!
\left[^{\underline{\hspace{3pt}}\,\Delta_3\;\underline{\hspace{3pt}}\,\Delta_2}_{\hspace{3pt}\,\Delta_4\;\hspace{3pt}\, \Delta_1} \right]\!(x)
    &=&
x^{\Delta - \underline{\hspace{3pt}}\,\Delta_2 - \Delta_1 }
\sum_{k\in \mathbb{N}- \frac{1}{2}}
\hskip -5pt
x^k\,
F^k_{c, \Delta}\!
\left[^{\underline{\hspace{3pt}}\,\Delta_3 \;\underline{\hspace{3pt}}\,\Delta_2}_{\hspace{3pt}\,\Delta_4\;\hspace{3pt}\, \Delta_1} \right],
\end{eqnarray}
superconformal blocks are determined by their
 coefficients:
\begin{eqnarray}
\label{block:definition} && F^{f}_{c, \Delta}\!
\left[^{\underline{\hspace{3pt}}\,\Delta_3
\;\underline{\hspace{3pt}}\,\Delta_2}_{\hspace{3pt}\,\Delta_4
\;\hspace{3pt}\, \Delta_1} \right] \; \\
\nonumber && = \hspace*{-20pt}
\begin{array}[t]{c}
{\displaystyle\sum} \\[2pt]
{\scriptstyle |K|+|M| = |L|+|N| = f }
\end{array}
\hspace*{-20pt} \rho^{\Delta_4\ \Delta_3 \ \Delta}_{\infty \ \ 1 \
\ \ 0} (\nu_4, \underline{\hspace{4pt}}\,\nu_3 , \nu_{\Delta,KM} )
\ \left[B^{f}_{c, \Delta}\right]^{KM,LN}  \rho^{\Delta\ \Delta_2 \
\Delta_1}_{\infty \ \ 1 \ \ \ 0}
   (\nu_{\Delta,LN},  \underline{\hspace{4pt}}\,\nu_2 , \nu_1 ),
\end{eqnarray}
where {\small $\left[B^{f}_{c, \Delta}\right]^{KM,LN}$} is the
matrix inverse to the Gram matrix of superconformal Verma module at level $f$
with respect to the basis $\{\nu_{\Delta,LN}\}$,
$\underline{\hspace{4pt}}\,\Delta_i$ and
$\underline{\hspace{4pt}}\,\nu_i$ stand for $\Delta_i$ or
$*\Delta_i$, and $\nu_i$ or $*\nu_i$, respectively, and
$x^{\Delta -*\Delta_2 - \Delta_1} =x^{\Delta - \Delta_2 - \Delta_1-{1\over 2}}. $

It follows from (\ref{block:definition}) that the blocks' coefficients
are polynomials in the external weights $\Delta_i$
and rational functions of the intermediate weight $\Delta$ and the central charge $c.$ They
can be expressed  as a sum over the poles in~$\Delta:$
\begin{equation}
\label{first:expansion:Delta}
F^{f}_{c, \Delta}\!
\left[^{\underline{\hspace{3pt}}\,\Delta_3
\;\underline{\hspace{3pt}}\,\Delta_2}_{\hspace{3pt}\,\Delta_4
\;\hspace{3pt}\, \Delta_1} \right]
\; = \;
{\rm h}^{f}_{c,\Delta}\!
\left[^{\underline{\hspace{3pt}}\,\Delta_3
\;\underline{\hspace{3pt}}\,\Delta_2}_{\hspace{3pt}\,\Delta_4
\;\hspace{3pt}\, \Delta_1} \right]
+
\begin{array}[t]{c}
{\displaystyle\sum} \\[-6pt]
{\scriptscriptstyle
1 < rs \leq 2{f}}
\\[-8pt]
{\scriptscriptstyle
r + s\in 2{\mathbb N}
}
\end{array}
\frac{
{\mathcal R}^{{f}}_{c,\,rs}\!
\left[^{\underline{\hspace{3pt}}\,\Delta_3
\;\underline{\hspace{3pt}}\,\Delta_2}_{\hspace{3pt}\,\Delta_4
\;\hspace{3pt}\, \Delta_1} \right]
}
{
\Delta-\Delta_{rs}(c)
}\,,
\end{equation}
with $\Delta_{rs}(c)$ given by Kac determinant formula for NS Verma modules:
\begin{eqnarray}
\label{delta:rs}
\Delta_{rs}(c)
& = &
-\frac{rs-1}{4} + \frac{1-r^2}{8}b^2 + \frac{1-s^2}{8}\frac{1}{b^2}\,,\hskip 10mm
c=\frac{3}{2} +3\left(b+{1\over b}\right)^2.
\end{eqnarray}
It was shown in \cite{Hadasz:2006sb} that
 the residue at $\Delta = \Delta_{rs}$ takes the form
\begin{eqnarray}
\label{res:evenf} {\mathcal R}^{m}_{c,\,rs}\!
\left[^{\underline{\hspace{3pt}}\,\Delta_3
\;\underline{\hspace{3pt}}\,\Delta_2}_{\hspace{3pt}\,\Delta_4
\;\hspace{3pt}\, \Delta_1} \right] &=&
S_{rs}(\underline{\hspace{3pt}}\Delta_3)
A_{rs}^c\hspace{-3pt}\left[^{\underline{\hspace{3pt}}\,\Delta_3
\;\underline{\hspace{3pt}}\,\Delta_2}_{\hspace{3pt}\,\Delta_4
\;\hspace{3pt}\, \Delta_1} \right] F^{m-\frac{rs}{2}}_{c,
\Delta_{rs} + \frac{rs}{2}}\!
\left[^{\underline{\hspace{3pt}}\,\Delta_3
\;\underline{\hspace{3pt}}\,\Delta_2}_{\hspace{3pt}\,\Delta_4
\;\hspace{3pt}\, \Delta_1} \right]
\end{eqnarray}
for $m \in {\mathbb N}\cup \{0\}$ and
\begin{eqnarray}
\label{res:oddf} {\mathcal R}^{k}_{c,\,rs}\!
\left[^{\underline{\hspace{3pt}}\,\Delta_3
\;\underline{\hspace{3pt}}\,\Delta_2}_{\hspace{3pt}\,\Delta_4
\;\hspace{3pt}\, \Delta_1} \right] &=&
S_{rs}(\underline{\hspace{3pt}}\Delta_3)A_{rs}^c\hspace{-3pt}\left[^{\widetilde{\underline{\hspace{3pt}}\,\Delta_3}
\;\widetilde{\underline{\hspace{3pt}}\,\Delta_2}}_{\hspace{3pt}\,\Delta_4
\;\hspace{3pt}\, \Delta_1} \right] F^{k-\frac{rs}{2}}_{c,
\Delta_{rs} + \frac{rs}{2}}\!
\left[^{\underline{\hspace{3pt}}\,\Delta_3
\;\underline{\hspace{3pt}}\,\Delta_2}_{\hspace{3pt}\,\Delta_4
\;\hspace{3pt}\, \Delta_1} \right]
\end{eqnarray}
for $k \in {\mathbb N}- \frac12.$ Here
\(
\widetilde{*\Delta}=\Delta,
\)
\(
\widetilde{\Delta}=*\Delta,
\)
\(
S_{rs}(\Delta)=1,
\)
\(
S_{rs}(*\Delta)=(-1)^{rs}
\)
and
\begin{eqnarray*}
 A_{rs}^c\hspace{-3pt}\left[^{{\underline{\hspace{3pt}}\,\Delta_3}
\;{\underline{\hspace{3pt}}\,\Delta_2}}_{\hspace{3pt}\,\Delta_4
\;\hspace{3pt}\, \Delta_1} \right] =
A_{rs}(c)\,
P^{rs}_{c}\!\left[^{\underline{\hspace{3pt}}\,\Delta_3}_{\hspace{5pt}\Delta_4}\right]
P^{rs}_{c}\!\left[^{\underline{\hspace{3pt}}\,\Delta_2}_{\hspace{5pt}\Delta_1}\right],
\end{eqnarray*}
where
\begin{equation}
\label{P:1} P^{rs}_{c}\!\left[^{\Delta_2}_{\Delta_1}\right] \; =
\; \prod_{p=1-r}^{r-1} \prod_{q=1-s}^{s-1}
\left(\frac{2a_1-2a_2 -p b - q b^{-1}}{2\sqrt2}\right) \left(\frac{2a_1 + 2a_2 +p b + q b^{-1}}{2\sqrt2}\right)
\end{equation}
with $p+q -(r+s) \in 4{\mathbb Z} + 2,$
\begin{equation}
\label{P:2}
P^{rs}_{c}\!\left[^{*\Delta_2}_{\hspace{4pt}\Delta_1}\right] \; =
\; \prod_{p=1-r}^{r-1} \prod_{q=1-s}^{s-1}
\left(\frac{2a_1-2a_2 -p b - q b^{-1}}{2\sqrt2}\right) \left(\frac{2a_1 + 2a_2 +p b + q b^{-1}}{2\sqrt2}\right)
\end{equation}
with $p+q -(r+s) \in 4{\mathbb Z}$ and
\begin{equation}
\label{A:rs:2}
A_{rs}(c)
\; = \;
\frac12
\prod_{p=1-r}^r
\prod_{q=1-s}^s
\left(\frac{1}{\sqrt{2}}\left(p b + \frac{q}{b}\right)\right)^{-1}
\hskip -3mm,
\hskip .5cm
p+q \in 2{\mathbb Z}, \; (p,q) \neq (0,0),(r,s).
\end{equation}

\section{Supersymmetric Liouville theory and classical limit of superconformal blocks}
\setcounter{equation}{0}

Within a path integral approach  the $N=1$ super-Liouville theory
is defined by the action:
\begin{equation}
\label{L:SLFT}
{\cal S}_{\rm\scriptscriptstyle SLFT}=
\int\!d^2z
\left(\frac{1}{2\pi}\left|\partial\phi\right|^2
+
\frac{1}{2\pi}\left(\psi\bar\partial\psi + \bar\psi\partial\bar\psi\right)
+
2i\mu b^2\bar\psi\psi {\rm e}^{b\phi}
+
2\pi b^2\mu^2{\rm e}^{2b\phi}\right).
\end{equation}
Each super-primary field $V_a$ with conformal dimension $
\Delta_a=\bar\Delta_a={a(Q-a)\over 2} $ and all its descendant
Virasoro primaries are represented by exponentials:
$$
\begin{array}{lllllllllll}
V_a &=& {\rm e}^{a\phi},
\\[5pt]
\Lambda _a &=& \left[S_{-1/2},V_a\right]
=
-ia\psi{\rm e}^{a\phi},
\\[5pt]
\bar\Lambda _a &=& \left[\bar S_{-1/2},V_a\right]
=
-i a\bar \psi{\rm
e}^{a \phi},
\\[5pt]
\widetilde V_a &=& \left\{S_{-1/2},\left[\bar S_{-1/2},V_a\right]
\right\}
=
a^2\psi\bar\psi{\rm e}^{a\phi} -2i\pi\mu b a {\rm e}^{(a
+b)\phi}.
\end{array}
$$
One has for instance
\begin{equation}
\label{cor} \langle V_{4} V_{3}\tilde V_{2}V_{1} \rangle=
\int\limits {\cal D}\phi {\cal D}\psi {\cal D}\bar\psi\;
{\rm e}^{-{\cal S}_{\rm\scriptscriptstyle SLFT}[\phi,\psi]}
{\rm e}^{a_4\phi} {\rm e}^{a_3\phi}
\left( a^2_2\psi\bar\psi{\rm e}^{a_2\phi} -2i\pi\mu b a_2 {\rm e}^{(a_2 +b)\phi} \right)
{\rm e}^{a_1\phi}.
\end{equation}
In order to analyze the classical limit ($b\to 0$, $2\pi \mu b^2 \to m={\rm const}$)
of this correlator
one may  integrate  fermions out.
Since the integration  is gaussian and the operator
${\rm e}^{b\phi}$ is light, one can expect
in the case of
heavy weights
\[
a = \textstyle \frac{Q}{2}\left(1-\lambda\right),
\hskip 5mm
ba\to {1-\lambda\over 2},
\hskip 5mm
2{b^2}\Delta\to \delta={1-\lambda^2\over 4},
\]
the following
asymptotic behavior
$$
\langle V_{4} V_{3}\tilde V_{2}V_{1}\rangle
\sim
{\textstyle {1\over b^2}}\,
{\rm e}^{-{1\over 2 b^2}S_{\rm\scriptscriptstyle cl}[\delta_4,\delta_3,\delta_2,\delta_1]},
$$
where $S_{\rm\scriptscriptstyle cl}[\delta_4,\delta_3,\delta_2,\delta_1]$ is the bosonic  Liouville
action\footnote{The factor ${1\over 2}$ in front of the Liouville classical action cames
from different parameterizations of the central charge in the NS and in the bosonic Liouville theory.}
$$
S[\phi]=
\frac{1}{2\pi}
\int\!d^2z
\left(\left|\partial\phi\right|^2
+
 m^2{\rm e}^{2\phi}\right)
$$
calculated on the classical configuration $\varphi$ satisfying the Liouville equation
$$
\partial\bar\partial \varphi - {m^2}{\rm e}^{2\phi}=\sum\limits_{i=1}^4 {1-\lambda_i\over 4}\delta(z-z_i).
$$

On the other hand in $N=1$ supersymmetric Liouville theory the
correlator $\langle V_{4}V_{3}\tilde V_{2}V_{1} \rangle $
can be expressed as an integral over the spectrum. In the case of
standard locations $z_4=\infty, z_3=1, z_2=x, z_1=0$ one has
\begin{equation}
\label{ql}
\langle V_{4} V_{3}\tilde V_{2}V_{1} \rangle
\;\ =
\int\limits_{\frac{Q}{2} + i{\mathbb R}_+}
\hskip -10pt
\frac{da}{2\pi i}
\Big(
C_{43s} \tilde C_{s21}
\left|\mathcal{F}_{\Delta}^{1}\!\left[^{\Delta_3 \;*\Delta_2}_{\Delta_4 \;\hspace{4.5pt}\Delta_1} \right]\!(x)\right|^2
         +
\tilde{C}_{43a} {C}_{a 21}
\left| \mathcal{F}_{\Delta}^{1\over 2}\!\left[^{\Delta_3\;*\Delta_2}_{\Delta_4\;\hspace{4.5pt}\Delta_1} \right]\!(x) \right|^2
\Big),
\end{equation}
where $C$ and $\tilde C$ are the the two independent supersymmetric Liouville structure constants
\[
C_{a 21}=
\langle
V_{a}(\infty,\infty)V_{a_2}(1,1)V_{a_1}(0,0)
\rangle ,
\hskip 1cm
\tilde C_{a 21}
=\langle
V_{a}(\infty,\infty)\tilde V_{a_2}(1,1)V_{a_1}(0,0)
\rangle.
\]
As in the case of 4-point functions the path integral representation yields
the asymptotic behavior
\begin{equation}
\label{3as}
\begin{array}{llll}
C_{a 21}
&\sim &
{\rm e}^{-{1\over 2 b^2}S_{\rm\scriptscriptstyle cl}[\delta,\delta_2,\delta_1]},
\\
\tilde C_{a 21}
& \sim &
{\textstyle{1\over b^2}}\,
{\rm e}^{-{1\over 2 b^2}S_{\rm\scriptscriptstyle cl}[\delta,\delta_2,\delta_1]},
\end{array}
\end{equation}
where $S_{\rm\scriptscriptstyle cl}[\delta,\delta_2,\delta_1]$ is the 3-point classical bosonic Liouville action.

Following the reasoning of \cite{Zamolodchikov:3,Zamolodchikov:1995aa} one can apply the path integral arguments
 to
 the   correlator (\ref{cor}) projected on the even and on the odd subspaces of
the $\Delta$ superconformal family of intermediate states.
This leads to the following $b\to 0$ asymptotic behavior:
\begin{equation}
\label{4as}
\begin{array}{lll}
\langle V_{4} V_{3}\mid_{\Delta}^{\rm \scriptscriptstyle even}\tilde V_{2}V_{1}\rangle
& \sim &
{\textstyle {1\over b^2}}\,
{\rm e}^{-{1\over 2 b^2}S_{\rm\scriptscriptstyle cl}[\delta_4,\delta_3,\delta_2,\delta_1|\delta]},
\\[5pt]
\langle V_{4} V_{3}\mid_{\Delta}^{\rm \scriptscriptstyle odd}\tilde V_{2}V_{1}\rangle
& \sim &
{\textstyle {1\over b^2}}\,
{\rm e}^{-{1\over 2 b^2}S_{\rm\scriptscriptstyle cl}[\delta_4,\delta_3,\delta_2,\delta_1|\delta]},
\end{array}
\end{equation}
where the ``$\Delta$-projected'' classical action is given by
$$
S_{\rm\scriptscriptstyle cl}[\delta_4,\delta_3,\delta_2,\delta_1|\delta]
=
S_{\rm\scriptscriptstyle cl}[\delta_4,\delta_3,\delta]
+S_{\rm\scriptscriptstyle cl}[\delta,\delta_2,\delta_1]
-
 f_{\delta}
\!\left[_{\delta_{4}\;\delta_{1}}^{\delta_{3}\;\delta_{2}}\right]
\!(x)
- \bar f_{\delta}
\!\left[_{\delta_{4}\;\delta_{1}}^{\delta_{3}\;\delta_{2}}\right]
\!(\bar x)
$$
and
\(
f_{\delta}\!\left[_{\delta_{4}\;\delta_{1}}^{\delta_{3}\;\delta_{2}}\right]\!(x)
\)
is the classical conformal block, defined in terms of
the $\tilde Q\to \infty$ limit of the quantum  conformal block
in the Virasoro $c=1+6\tilde Q^2$ CFT:
\begin{equation}
\label{defccb}
{\cal F}_{\!1+6\tilde Q^2,\Delta}
\!\left[_{\Delta_{4}\;\Delta_{1}}^{\Delta_{3}\;\Delta_{2}}\right]\!(x)
\; \stackrel{\tilde Q \to \infty}{\sim} \;
\exp \left\{
\tilde Q^2\,f_{\delta}
\!\left[_{\delta_{4}\;\delta_{1}}^{\delta_{3}\;\delta_{2}}\right]\!(x)
\right\}.
\end{equation}
Equations (\ref{ql}) -- (\ref{4as}) then
imply:
\begin{eqnarray}
\label{claslim1}
\mathcal{F}_{\Delta}^{1}\!
\left[^{\Delta_3 \;*\Delta_2}_{\Delta_4 \;\hspace{4.5pt}\Delta_1} \right]\!(x)
&\sim &
{\rm e}^{{1\over 2 b^2} f_{\delta}
\!\left[_{\delta_{4}\;\delta_{1}}^{\delta_{3}\;\delta_{2}}\right](x)},
\hskip 1cm
\mathcal{F}_{\Delta}^{1\over 2}\!
\left[^{\Delta_3\;*\Delta_2}_{\Delta_4\;\hspace{4.5pt}\Delta_1} \right]\!(x)
\;\sim \;
{\rm e}^{{1\over 2b^2} f_{\delta}
\!\left[_{\delta_{4}\;\delta_{1}}^{\delta_{3}\;\delta_{2}}\right](x)}.
\end{eqnarray}
Using representations analogous to (\ref{ql}) and the same
reasoning for the other 4-point correlators of primary fields $V_a,
\Lambda_a, \bar \Lambda_a, \tilde V_a$ one gets
\begin{equation}
\label{claslim2}
\begin{array}{lllllllll}
\mathcal{F}_{\Delta}^{1}\!
\left[^{\Delta_3\;\Delta_2}_{\Delta_4\;\Delta_1} \right]\!(x)
& \sim &
{\rm e}^{{1\over 2b^2} f_{\delta}
\!\left[_{\delta_{4}\;\delta_{1}}^{\delta_{3}\;\delta_{2}}\right](x)},
&&
\mathcal{F}_{\Delta}^{1\over 2}\!
\left[^{\Delta_3\;\Delta_2}_{\Delta_4\;\Delta_1} \right]\!(x)
& \sim &
b^2\,
{\rm e}^{{1\over 2b^2} f_{\delta}
\!\left[_{\delta_{4}\;\delta_{1}}^{\delta_{3}\;\delta_{2}}\right](x)},
\\[6pt]
\mathcal{F}_{\Delta}^{1}\!
\left[^{*\Delta_3\;*\Delta_2}_{\hspace{4.5pt}\Delta_4\;\hspace{4.5pt}\Delta_1} \right]\!(x)
& \sim &
{\rm e}^{{1\over 2b^2} f_{\delta}
\!\left[_{\delta_{4}\;\delta_{1}}^{\delta_{3}\;\delta_{2}}\right](x)},
&&
\mathcal{F}_{\Delta}^{1\over 2}\!
\left[^{*\Delta_3\;*\Delta_2}_{\hspace{4.5pt}\Delta_4\;\hspace{4.5pt}\Delta_1} \right]\!(x)
& \sim &
{\textstyle {1\over b^2}}\,
{\rm e}^{{1\over 2b^2} f_{\delta}
\!\left[_{\delta_{4}\;\delta_{1}}^{\delta_{3}\;\delta_{2}}\right](x)}.
\end{array}
\end{equation}
The properties of classical conformal block relevant for the
the elliptic recurrence relations
where already derived by Al.~B.~Zamolodchikov
in the Virasoro CFT \cite{Zamolodchikov:3}.
In the next two sections we shall nevertheless present  a step by step derivation of
these properties  in NS SCFT. This can be seen as a nontrivial consistency
check of heuristic path integral arguments of this section.

\section{Large $\Delta_p$ vs.\ classical asymptotic of superconformal blocks}
\setcounter{equation}{0}

As in the bosonic case the first step in the derivation of the elliptic recurrence
is to find the large $\Delta$ asymptotic of the conformal block.
The method of calculations proposed in \cite{Zamolodchikov:3} is based
on the observation that the full dependence of the first two terms in the large $\Delta$ expansion
on the variables $\Delta_i, c$
can be read off from the first two terms of the ${1\over \delta}$ expansion
of the classical block. While in the case of even NS blocks the reasoning is essentially the same
as in \cite{Zamolodchikov:3}, the odd case is slightly more complicated.

Let us first note that on each level of the NS Verma module the
determinant of the Gram matrix is proportional to $\Delta$
and each matrix element of its inverse contains a
factor $\Delta^{-1}.$ On the other hand it follows from the
properties of the 3-point superconformal blocks
\cite{Hadasz:2006sb} that in a generic case $\rho^{\Delta_4\
\Delta_3 \ \Delta}_{\infty \ \ 1 \ \ \ 0} (\nu_4, \nu_3 ,
\nu_{\Delta,KM} )$ does not contain the factor $\Delta$, while for
all odd levels $|K|\in \mathbb{N}-{1\over 2}$, $\rho^{\Delta_4\
\Delta_3 \ \Delta}_{\infty \ \ 1 \ \ \ 0} (\nu_4, *\nu_3 ,
\nu_{\Delta,KM} )$ is proportional to $\Delta$. Since
the power series defining the odd
blocks (\ref{odd_block_def}) do not contain zeroth order terms it
follows from the definition (\ref{block:definition}) that the
functions
$$
\mathcal{G}_{\Delta}^{1\over 2}\!
\left[^{\underline{\hspace{3pt}}\,\Delta_3\;\underline{\hspace{3pt}}\,\Delta_2}_{\hspace{3pt}\,\Delta_4\;\hspace{3pt}\, \Delta_1} \right]\!(x)
\; = \;
\ln
\mathcal{F}_{\Delta}^{1\over 2}\!
\left[^{\underline{\hspace{3pt}}\,\Delta_3\;\underline{\hspace{3pt}}\,\Delta_2}_{\hspace{3pt}\,\Delta_4\;\hspace{3pt}\, \Delta_1} \right]\!(x)
$$
admit power series expansions of the form
\begin{eqnarray*}
\mathcal{G}_{\Delta}^{1\over 2}\!
\left[^{\Delta_3
\;\Delta_2}_{\Delta_4
\;\Delta_1} \right]\!(x)
         &=&
(\Delta - \Delta_2 - \Delta_1)\ln x -\ln \Delta + \sum\limits_{i=0}^\infty
G_n x^n,
\\
\mathcal{G}_{\Delta}^{1\over 2}\!
         \left[^{\Delta_3
\;*\Delta_2}_{\Delta_4
\;\hspace{4.5pt}\Delta_1} \right]\!(x)
         &=&
(\Delta - *\Delta_2 - \Delta_1)\ln x + \sum\limits_{i=0}^\infty
G^*_n x^n,
\\
\mathcal{G}_{\Delta}^{1\over 2}\!
         \left[^{*\Delta_3
\;*\Delta_2}_{\hspace{4.5pt}\Delta_4
\;\hspace{4.5pt}\Delta_1} \right]\!(x)
         &=&
(\Delta - *\Delta_2 - \Delta_1)\ln x + \ln \Delta+\sum\limits_{i=0}^\infty
G^{**}_n x^n,
\end{eqnarray*}
where the coefficients $G_n, G^*_n, G^{**}_n $ are rational functions of $\Delta, \Delta_i, c$.

We shall consider  the first case  in more detail. One has:
\begin{eqnarray*}
G_n&=&{P_n(\Delta,\Delta_i,c)\over Q_n(\Delta, c)},
\end{eqnarray*}
where  $P_n(\Delta,\Delta_i,c)$ and $ Q_n(\Delta, c)$ are polynomials in all their arguments. The existence of semiclassical limit
(\ref{claslim2}) implies that the maximal homogeneous degree of $P_n(\Delta,\Delta_i,c),$
$$
P^{N_n+1}_n(\Delta,\Delta_i,c)= \Delta^{N_n} \left(A_n \Delta + \sum\limits_{i=1}^4
B^i_n \Delta_i + C_n c\right) + \Delta^{N_n-1}\left( \dots \right.,
$$
is greater by 1 than the maximal homogeneous  degree of $Q_n(\Delta, c),$
$$
Q^{N_n}_n(\Delta, c) = a_n\Delta^{N_n} + b_n \Delta^{N_n-1} c + c_n \Delta^{N_n-2} c^2 +\dots
$$
The coefficients $A_n,\ldots c_n$ are by definition independent of $c,\Delta$ and $\Delta_i.$

It follows from (\ref{block:definition}) and the Kac determinant formula for NS supermodules
that $a_n\neq 0$,
hence
$$
f_{\delta}^n
\!\left[_{\delta_{4}\;\delta_{1}}^{\delta_{3}\;\delta_{2}}\right]
={\delta^{N_n} \left(A_n \delta + \sum\limits_{i=1}^4
B^i_n \delta_i + 6 C_n \right) + \delta^{N_n-1}\left( \dots \right.\over
 a_n\delta^{N_n} +  6 b_n \delta^{N_n-1} + 36 c_n \delta^{N_n-2} +\dots}.
$$
Expanding in reciprocal powers of $\delta$ one gets
$$
f_{\delta}^n
\!\left[_{\delta_{4}\;\delta_{1}}^{\delta_{3}\;\delta_{2}}\right]
={A_n\over a_n}  \delta
 + \sum\limits_{i=1}^4{ B^i_n  \over a_n}\delta_i + {6( C_n +A_n b_n)\over a_n}
 + {\cal O}\left({1\over \delta}\right).
$$
On the other hand the ${1\over \Delta}$ expansion of
$G_n$ takes the form
\begin{equation}
\label{G_eq}
 G_n={A_n\over a_n}  \Delta
 + \sum\limits_{i=1}^4 { B^i_n \over a_n} \Delta_i +  {C_n  + A_n b_n \over a_n} c+
 {D_n\over a_n} +{\cal O} \left({1\over \Delta}\right),
\end{equation}
where $D_n$ is the coefficient in front of $\Delta^{N_n}$ in the polynomial $P_n.$

\section{Null vector decoupling equation}
\setcounter{equation}{0}

We  consider 5-point correlators of primary fields $V_{i} =
V_{a_i}(z_i, \bar{z}_i) $ or $\Lambda_{i} = \Lambda_{a_i}(z_i,
\bar{z}_i) $
 in the limit $a_5 \to -b.$ The field $V_{-b}\left(z_5,\bar z_5\right)$
is degenerate and satisfies the null vector decoupling equation:
\begin{eqnarray*}
V_0(z_5,\bar z_5) \equiv \left(L_{-1}S_{-\frac12} + b^2
S_{-\frac32}\right)V_{-b}\left(z_5,\bar z_5\right) = 0.
\end{eqnarray*}
Applying to correlators $\langle V_{4}
\Lambda_{3} V_{0} V_{2}V_{1}\rangle$, $\langle V_{4} V_{3} V_{0}
\Lambda_{2}V_{1}\rangle$, $\langle V_{4} V_{3} V_{0}
V_{2}\Lambda_{1}\rangle$, $\langle V_{4} \Lambda_{3} V_{0}
\Lambda_{2}\Lambda_{1}\rangle$
the conformal Ward identities
one can obtain differential
equations ($ z_4 \to \infty  $):
\begin{eqnarray}
\nonumber
&& \hskip -1cm \left[
\partial^2_{5} + b^2 \left(
\frac{1}{z_{53}} \partial_{3} +
\frac{1}{z_{52}} \partial_{2} +
\frac{1}{z_{51}} \partial_{1} +
\frac{2\Delta_1}{z_{51}^2} \right) \right]
\langle V_{4} V_{3} V_5 V_2 V_1\rangle
\\[3pt]
\nonumber
& = &
\left( \partial_{z_5} - \frac{b^2}{z_{52}} \right)
\langle V_{ 4} V_3 \Lambda_5 \Lambda_{2} V_1\rangle - \left(
\partial_{z_5} - \frac{b^2}{z_{53}} \right)
 \langle V_4 \Lambda_{3} \Lambda_5 V_2V_1\rangle
\\ [3pt]
\nonumber
&+& b^2
\left( \frac{1}{z_{53}} - \frac{1}{z_{52}} \right)
\langle V_4 \Lambda_{3} V_5 \Lambda_{2}V_1 \rangle,
\\[10pt]
\label{5:point:after:limit}
&& \hskip -1cm
b^2 \left[
\left( \frac{1}{z_{15}} + \frac{1}{z_{52}} \right) \partial_{2} + \frac{2\Delta_2}{z_{52}^2} \right]
\langle V_4 V_3 V_5 V_2V_1\rangle
\\ [3pt]
\nonumber
 & = &
b^2 \left( \frac{1}{z_{35}} + \frac{1}{z_{51}} \right)
 \langle V_4 \Lambda_{3} V_5 \Lambda_{2}V_1\rangle
-\left( \partial_{z_5} + \frac{b^2}{z_{15}} \right)
\langle V_4 V_3 \Lambda_5 \Lambda_{2}V_1\rangle,
\\[10pt]
\nonumber
&& \hskip -1cm
 b^2 \left[
\left( \frac{1}{z_{15}} + \frac{1}{z_{53}} \right)
\partial_{3} + \frac{2\Delta_3}{z_{53}^2} \right]
\langle V_4 V_3 V_5 V_2V_1\rangle
\\ [3pt]
\nonumber
 & = &
- b^2 \left( \frac{1}{z_{25}} + \frac{1}{z_{51}} \right)
\langle V_4 \Lambda_{3} V_5 \Lambda_{2}V_1\rangle
+ \left( \partial_{z_5} + \frac{b^2}{z_{15}} \right)
\langle V_4 \Lambda_{3} \Lambda_5 V_2V_1\rangle .
\end{eqnarray}
Adding the second equation to the first one and subtracting from
the result the third equation   we obtain:
 \begin{eqnarray}
&& \hskip -1cm \left[
\partial^2_{z_5}\! + b^2 \left(
\frac{1}{z_{51}} \partial_{1} +
\left( \frac{1}{z_{15}} + \frac{2}{z_{52}} \right) \partial_{2}
+\left( \frac{1}{z_{15}} + \frac{2}{z_{53}} \right) \partial_{3}
+ \frac{2\Delta_1}{z_{51}^2}
+ \frac{2\Delta_2}{z_{52}^2}
+ \frac{2\Delta_3}{z_{53}^2}
 \right) \right]
\langle V_4 V_3 V_{5} V_2V_1\rangle
\nonumber
\\[3pt]
\label{suma}
 &=&
b^2
 \left( \frac{1}{z_{51}} - \frac{1}{z_{52}} \right)
\langle V_4 V_3 \Lambda_5
\Lambda_{2} V_1\rangle - b^2\left(
\frac{1}{z_{51}} - \frac{1}{z_{53}} \right)
 \langle V_4 \Lambda_{3} \Lambda_5 V_2V_1\rangle
\end{eqnarray}
 Since $V_{-b}$  and  $\Lambda_{-b}$ are ``light'' fields, in the classical limit
all  the correlators in (\ref{suma})  have the form
 $$
 \chi(z_5)\,
{\rm e}^{-\frac{1}{2 b^2}S_{\rm\scriptscriptstyle
cl}[\delta_4,\delta_3,\delta_2,\delta_1]}\ .
$$
Therefore, for $b \to 0$ and $ \Delta_1, \Delta_2, \Delta_3, \Delta_4$ of order $b^{-2},$
 we have:
\begin{eqnarray*}
\partial_{1},\; \partial_{2},\; \partial_{3}  = {\cal O}\left(b^{-2}\right),
\qquad  \Delta_5,\; \partial_{z_5} =  {\cal O}\left(1\right).
\end{eqnarray*}
Keeping only the leading terms  in (\ref{suma}) we thus get the
closed equation for the classical limit of $ \langle V_4 V_3 V_5
V_2V_1\rangle$. In the standard locations $z_1= 0, z_3 = 1,$  $z_5
= z,\ z_2 = x,$ it takes the form:
\begin{eqnarray}
\label{eigen:2}
\nonumber
\Bigg\lbrace \partial^2_z
&+& 2b^2\left[
\frac{\Delta_4-\Delta_3-\Delta_2-\Delta_1}{z(z-1)}
+
\frac{\Delta_3}{(z-1)^2}
+
\frac{\Delta_2}{(z-x)^2}
+
\frac{\Delta_1}{z^2}
\right] \Bigg\rbrace
\langle V_4 V_3 V_5 V_2 V_1\rangle
\\[4pt]
&+& 2b^2\frac{x(x-1)}{z(z-1)(z-x)}
\frac{\partial}{\partial x} \,
\langle V_4 V_3 V_5 V_2 V_1\rangle
 =  0.
\end{eqnarray}

Let us consider the contribution  to this correlation function
 from an even subspace of a  single NS Verma module
${\cal V}_{\Delta}\!\!\otimes \overline{{\cal V}}_{\Delta}$.
In the classical limit one gets:
\begin{eqnarray*}
 \Big\langle
V_4\left(\infty\right) V_3\left(1,1\right) V_{-b}\left(z,\bar
z\right) \mid_{\Delta}^{\rm \scriptscriptstyle even}
V_2\left(x,\bar x\right) V_1\left(0,0\right) \Big\rangle \sim
\chi_\Delta(z)\ {\rm e}^{
\frac{1}{2b^2}f_{\delta\!}\left[^{\delta_3\:\delta_2}_{\delta_4\:\delta_1}\right]\!(x)},
\end{eqnarray*}
where \(
f_{\delta\!}\left[^{\delta_3\:\delta_2}_{\delta_4\:\delta_1}\right]\!(x)
\) is the classical conformal block (\ref{defccb}).
Substituting into (\ref{eigen:2}) one obtains a Fuchsian equation:
\begin{eqnarray}
\label{eigen:3}
\frac{d^2\chi_\Delta(z)}{dz^2}
+
\left(
\frac{\delta_4 - \delta_3 -\delta_2 - \delta_1}{z(z-1)}
+
\frac{\delta_1}{z^2} + \frac{\delta_2}{(z-x)^2} + \frac{\delta_3}{(z-1)^2}
\right)
\chi_\Delta(z)
&&
\\[10pt]
\nonumber
+
\frac{x(x-1){\cal C}(x)}{z(z-x)(z-1)}
\chi_\Delta(z)
& = &
0,
\end{eqnarray}
with the
accessory parameter ${\cal C}(x)$ given by:
\begin{eqnarray}
\label{accessory:parameter}
{\cal C}(x)
& = &
\frac{\partial}{\partial x}
f_{\delta}\!\!\left[^{\delta_3\:\delta_2}_{\delta_4\:\delta_1}\right]\!(x).
\end{eqnarray}

We shall now calculate the monodromy properties of $\chi_\Delta(z)$ along
the contour encircling the points $0$ and $x$. There are only
three conformal families in the OPE of the degenerate field
$V_{-b}$ with a super-primary field $V_{a}$:
\begin{eqnarray}
\nonumber \label{OPE:degenerate}  V_{-b}(z, \bar
z)V_{a}(0, 0)
 &=& C_{a_+,-b,a}\,(z \bar z)^{\frac{bQ}{2}(1+\lambda)}\,V_{a_+}(0,0) +
 C_{a_-,-b,a}\,(z \bar z)^{\frac{bQ}{2}(1-\lambda)}\,V_{a_-}(0,0)
\\ [6pt]
&+& \tilde C_{a,-b,a}\,(z \bar z)^{1+b^2}\, \frac{1}{(2
\Delta_{a})^2} \, \tilde V_{a}(0,0)
 \, + \, {\rm descendants},
\end{eqnarray}
where
$
a_{\pm} = a \pm b.
$

In the classical limit the third term  in (\ref{OPE:degenerate})
is sub-leading with respect to the first two. Hence  in the space
of solutions of (\ref{eigen:3}) there is a basis $\chi_{\Delta}^\pm(z)$
such that:
\begin{eqnarray}
\label{basis:continuation}
\chi_{\Delta}^\pm\left({\rm e}^{2\pi i}z\right)
& = &
-{\rm e}^{\pm i\pi \lambda}\,\chi_{\Delta}^\pm\left(z\right).
\end{eqnarray}
The problem of calculating ${\cal C}$ can then be formulated as follows:
adjust ${\cal C}$ in such a way that the equation
admits solutions with the monodromy around $0$ and $x$ given by
(\ref{basis:continuation}). This is exactly the  monodromy problem
obtained and solved in the context of Virasoro theory in
\cite{Zamolodchikov:2,Zamolodchikov:3}. Details of these
calculations are presented  in Appendix A.

Taking into account the ${1\over \delta}$ expansion of
the classical block
(\ref{clas_block}) and (\ref{G_eq})
one gets:
\begin{eqnarray}
\label{asymptoticG}
\nonumber
&& \hskip -1cm \mathcal{G}_{\Delta}^{1\over 2}\!
         \left[^{\Delta_3 \;\Delta_2}_{\Delta_4 \;\Delta_1} \right]\!(x)
         = -\ln \Delta
+\pi \tau \left( \Delta - \frac{c}{24} \right)
+ \left( \frac{c}{8} - \Delta_1 - \Delta_2 - \Delta_3 - \Delta_4\right) \, \ln{K^2(x)} \\
 [6pt] &&
+ \left( \frac{c}{24} - \Delta_2 - \Delta_3\right)  \, \ln(1-x)
+  \left( \frac{c}{24} - \Delta_1 - \ \Delta_2\right)  \, \ln(x) + f^{1\over 2}(x) +{\cal O}\left({1\over \Delta}\right),
\end{eqnarray}
where
\(
K(x)
\)
is the complete elliptic integral of the first kind,
\(
\tau \equiv \tau(x) = i\frac{K(1-x)}{K(x)}
\)
is the half-period ratio and
$f^{1\over 2}(x)$ is a function of $x$ specific for each type of block and independent of $\Delta_i$ and $c$.
One can obtain corresponding formulae for other types of blocks in a similar way.
The exact form of the functions $f^{1,\frac12}(x)$ can be derived from analytic expressions for $c={3\over 2}$ NS superconformal blocks
with external weights $\Delta_i= {1\over 8}$ \cite{hjs}.

\section{Elliptic recursion relations }
\setcounter{equation}{0}

The large $\Delta$ asymptotic suggests the following form of superconformal blocks:
 \begin{eqnarray}
\label{Hblock}
  \mathcal{F}^{1, \frac12}_{\Delta}\!
\left[^{\underline{\hspace{3pt}}\,\Delta_3
\;\underline{\hspace{3pt}}\,\Delta_2}_{\hspace{3pt}\,\Delta_4
\;\hspace{3pt}\, \Delta_1} \right]\!(x)
 & =&
(16q)^{\Delta - \frac{c-3/2}{24}}\ x^{\frac{c-3/2}{24} - \Delta_1 - \underline{\hspace{3pt}}\, \Delta_2} \
 (1- x)^{\frac{c-3/2}{24} - \underline{\hspace{3pt}}\,\Delta_2 - \underline{\hspace{3pt}}\,\Delta_3}\
\\
\nonumber
 & \times &\theta_3^{\frac{c - 3/2}{2}
- 4 (\Delta_1 + \underline{\hspace{3pt}}\,\Delta_2 +\underline{\hspace{3pt}}\,\Delta_3 + \Delta_4) }  \
 \mathcal{H}^{1, \frac12}_{\Delta}\! \left[^{\underline{\hspace{3pt}}\,\Delta_3
\;\underline{\hspace{3pt}}\,\Delta_2}_{\hspace{3pt}\,\Delta_4
\;\hspace{3pt}\, \Delta_1} \right]\!(x),
\end{eqnarray}
where
\(
q \equiv q(x) = {\rm e}^{i\pi\tau}
\)
is the elliptic nome. The
elliptic blocks
\(\mathcal{H}^{1, \frac12}_{\Delta}\! \left[^{\underline{\hspace{3pt}}\,\Delta_3
\;\underline{\hspace{3pt}}\,\Delta_2}_{\hspace{3pt}\,\Delta_4
\;\hspace{3pt}\, \Delta_1} \right]\!(x)
\)
have the same analytic structure as
superconformal ones:
\begin{eqnarray*}
\mathcal{H}^{1, \frac12}_{\Delta}\! \left[^{\underline{\hspace{3pt}}\,\Delta_3
\;\underline{\hspace{3pt}}\,\Delta_2}_{\hspace{3pt}\,\Delta_4
\;\hspace{3pt}\, \Delta_1} \right]\!(x)=
  g^{1,\frac12\,}_{\underline{\hspace{3pt}}\,\underline{\hspace{3pt}}}(x)
+ \sum_{m,n}
\frac{
h^{1, \frac12}_{mn}
\left[^{\underline{\hspace{3pt}}\,\Delta_3
\;\underline{\hspace{3pt}}\,\Delta_2}_{\hspace{3pt}\,\Delta_4
\;\hspace{3pt}\, \Delta_1} \right]\!(x)
}{
\Delta - \Delta_{mn}}.
\end{eqnarray*}
The functions $ g^{1,\frac12\,}_{\underline{\hspace{3pt}}\,\underline{\hspace{3pt}}}(x)$
depend on the type of block and are independent
of the external weights $\Delta_i$ and the central charge $c$.
They have no singularities in $\Delta$ and are directly related to the functions
$f^{1,\frac12}(x)$ in (\ref{asymptoticG}). For instance, in the case of the odd block
$\mathcal{F}_{\Delta}^{1\over 2}\!
         \left[^{\Delta_3
\;\Delta_2}_{\Delta_4
\;\Delta_1} \right]\!(x):
$
$$
\exp f^{1\over 2}(x)
\; = \;
(16 q)^{1\over 16} \, \left[ x(1-x)\right] ^{-{1\over 16}} \, \theta_3(q)^{-{3\over 4}} \,
g^{1\over 2}(x).
$$
The analytic form of these functions can be read off from the form of the elliptic blocks
related to the $c = \frac32$ conformal ones with $\Delta_i= \Delta_0 = \frac18$ \cite{hjs}:
\begin{equation}
\label{H0blocks1}
\begin{array}{rlllrlllllllllll}
 \mathcal{H}^{1}_{\Delta}\! \left[^{\Delta_0 \; \Delta_0}_{\Delta_0 \; \Delta_0} \right]\!(q)
&=& \theta_3(q^2),
&\hspace{10pt}&
 \mathcal{H}^{\frac12}_{\Delta}\! \left[^{\Delta_0 \; \Delta_0}_{\Delta_0 \; \Delta_0} \right]\!(q)
&=&\frac{\textstyle 1}{\textstyle\Delta}\, \theta_2(q^2),
\\  [6pt]
\mathcal{H}^{1}_{\Delta}\! \left[^{\Delta_0 \;*\Delta_0}_{\Delta_0 \;\hspace{4pt}\Delta_0} \right]\!(q)
&=& \theta_3(q^2),
&\hspace{10pt}&
 \mathcal{H}^{\frac12}_{\Delta}\! \left[^{\Delta_0 \;*\Delta_0}_{\Delta_0 \;\hspace{4pt}\Delta_0} \right]\!(q)
&=& \theta_2(q^2),
\end{array}
\end{equation}
\vspace*{-10pt}
\begin{eqnarray}
\label{H0blocks2}
\nonumber
\mathcal{H}^{1}_{\Delta}\! \left[^{*\Delta_0 \;*\Delta_0}_{\hspace{4pt} \Delta_0 \;\hspace{4pt}\Delta_0} \right]\!(q)
&=&
\theta_3(q^2) \left( 1 - \frac{q}{\Delta} \theta_3^{-1}(q) \frac{\partial}{\partial q} \theta_3(q) + \frac{\theta_2^{4}(q)}{4 \Delta} \right) ,
\\[-6pt]
\\[-6pt]
\nonumber
\mathcal{H}^{\frac12}_{\Delta}\! \left[^{*\Delta_0 \;*\Delta_0}_{\hspace{4pt} \Delta_0 \;\hspace{4pt}\Delta_0} \right] \!(q)
&=&
- \theta_2(q^2)  \left( \Delta - q\, \theta_3^{-1}(q) \frac{\partial}{\partial q} \theta_3(q) + \frac{\theta_2^{4}(q)}{4 } \right) .
\end{eqnarray}
Indeed the functions $ g^{1,\frac12\,}_{\underline{\hspace{3pt}}\,\underline{\hspace{3pt}}}(x)$
are just given by  terms regular for $\Delta\to 0:$
\begin{equation}
\label{gfunctions}
\begin{array}{rlllrllllllllllllll}
g^{1}(x)
&=& \theta_3(q^2),
&\qquad &g^{1\over 2}_{\Delta}(x)&=& 0,
\\ [4pt]
 g^{1}_*(x)
&=& \theta_3(q^2),
&\qquad &
 g^{\frac12}_{*} (x)
 &=& \theta_2(q^2),
\\
 g^{1}_{**}(x)
&=& \theta_3(q^2),
&\quad &
 g^{\frac12\,}_{**} (x)
&=& - \theta_2(q^2) \left( \Delta - q\, \theta_3^{-1}(q)\,
\textstyle\frac{\textstyle\partial}{\textstyle\partial q} \theta_3(q) + \frac{\textstyle\theta_2^{4}(q)}{\textstyle 4} \right).
\end{array}
\end{equation}
Taking into account the form of the residue at $\Delta_{mn}$
(equations (\ref{res:evenf}), (\ref{res:oddf}))
one gets the general elliptic recursion
relations:
\begin{eqnarray}
\nonumber
\mathcal{H}^{1,\frac12}_{\Delta}\!
\left[^{\underline{\hspace{3pt}}\,\Delta_3
\;\underline{\hspace{3pt}}\,\Delta_2}_{\hspace{3pt}\,\Delta_4
\;\hspace{3pt}\, \Delta_1} \right]\!(x)
&=&
 g^{1,\frac12\,}_{\underline{\hspace{3pt}}\,\underline{\hspace{3pt}}}(x)
\\
\label{Hrek}
&+&
\hspace{-15pt}\begin{array}[t]{c}
{\displaystyle\sum} \\[-6pt]
{\scriptscriptstyle
m,n>0}
\\[-8pt]
{\scriptscriptstyle
m,n\in 2{\mathbb N}
}
\end{array}
 (16q)^{\frac{mn}{2}}
\frac{A_{rs}^c\hspace{-3pt}\left[^{\underline{\hspace{3pt}}\,\Delta_3
\;\underline{\hspace{3pt}}\,\Delta_2}_{\hspace{3pt}\,\Delta_4
\;\hspace{3pt}\, \Delta_1} \right]}{\Delta - \Delta_{mn}} \,
\mathcal{H}^{1,\frac12}_{\Delta_{mn}+\frac{mn}{2}}\!
\left[^{\underline{\hspace{3pt}}\,\Delta_3
\;\underline{\hspace{3pt}}\,\Delta_2}_{\hspace{3pt}\,\Delta_4
\;\hspace{3pt}\, \Delta_1} \right]\!(x)
 \\
\nonumber&+&
\hspace{-15pt}\begin{array}[t]{c}
{\displaystyle\sum} \\[-6pt]
{\scriptscriptstyle
m,n>0}
\\[-8pt]
{\scriptscriptstyle
m,n\in 2{\mathbb N}+1
}
\end{array}(16q)^{\frac{mn}{2}}
\frac{S_{rs}(\underline{\hspace{3pt}}\Delta_3)A_{rs}^c\hspace{-3pt}\left[^{\widetilde{\underline{\hspace{3pt}}\,\Delta_3}
\;\widetilde{\underline{\hspace{3pt}}\,\Delta_2}}_{\hspace{3pt}\,\Delta_4
\;\hspace{3pt}\, \Delta_1} \right]}{\Delta - \Delta_{mn}} \,
\mathcal{H}^{\frac12,1}_{\Delta_{mn}+\frac{mn}{2}}\!
\left[^{\underline{\hspace{3pt}}\,\Delta_3
\;\underline{\hspace{3pt}}\,\Delta_2}_{\hspace{3pt}\,\Delta_4
\;\hspace{3pt}\, \Delta_1} \right]\!(x).
\end{eqnarray}
Formula (\ref{Hrek}) is the main result of the present paper.

As a nontrivial consistency check of (\ref{Hrek}) one can verify
 that each pair of elliptic blocks in (\ref{H0blocks1}), (\ref{H0blocks2}),
 satisfy recursion relations (\ref{Hrek}) with the corresponding
 functions (\ref{gfunctions}).
This is done in Appendix B.

\setcounter{equation}{0}
\section*{Acknowledgements}
The work  was partially supported by the Polish State Research
Committee (KBN) grant no. N N202 0859 33.

\vskip 1mm
\noindent
The research of L.H.\ was supported by the Alexander von Humboldt Foundation.

\vskip 1mm
\noindent
P.S.\ is grateful to the faculty of the Institute of Theoretical Physics, University of Wroc\l{}aw,
for the hospitality.

\section*{Appendix A}
\setcounter{equation}{0}
\renewcommand{\theequation}{A.\arabic{equation}}

Consider the equation
\begin{eqnarray}
\label{class:1}
\frac{d^2\chi(z)}{dz^2}
+
U(z) \chi (z)
+
\frac{x(x-1){\cal C}(x)}{z(z-x)(z-1)}
\chi(z)
& = &
0,
\end{eqnarray}
with
\begin{equation}
\label{U}
U(z) = \frac14\left(
\frac{\lambda_1^2 + \lambda_2^2 + \lambda_3^2 -\lambda_4^2 - 2}{z(z-1)}
+
\frac{1-\lambda_1^2}{z^2} + \frac{1-\lambda_2^2}{(z-x)^2} + \frac{1-\lambda_3^2}{(z-1)^2}
\right).
\end{equation}
We want to choose ${\cal C}(x)$ such that (\ref{class:1}) admits a pair of solutions $\chi^{\pm}(z)$ satisfying
the monodromy condition
\begin{equation}
\label{monodromy:condition}
\chi^{\pm}\left({\rm e}^{2\pi i}z\right) = -{\rm e}^{\pm i\pi\lambda}\chi^{\pm}(z),
\end{equation}
where
\(
\chi\left({\rm e}^{2\pi i}z\right)
\)
denotes a function analytically continued in $z$ along a closed path encircling points $z = 0$ and $z = x.$

Following \cite{Zamolodchikov:3} we perform an elliptic change of variables:
\begin{equation}
\label{elliptic:substitution}
\xi(z) = \frac12\int\limits^{\frac{z}{x}}
\frac{dt}{\sqrt{t(1-t)(1-xt)}},
\qquad
\tilde\chi(\xi) = \left(\frac{dz(\xi)}{d\xi}\right)^{-\frac12}\chi\Big(z(\xi)\Big).
\end{equation}
This gives
\begin{eqnarray}
\label{second:derivative}
\frac{d^2}{dz^2} \chi(z)
& = &
  -\frac12 \left(\xi'\right)^{-\frac12}\left\{\xi(z),z\right\}\tilde\chi(\xi)
+
\left(\xi'\right)^{+\frac32}\left.\frac{d^2\tilde\chi(\xi)}{d\xi^2}\right|_{\xi = \xi(z)},
\end{eqnarray}
where
\(
\left\{\xi(z),z\right\}
\)
is the Schwarzian derivative  of the map (\ref{elliptic:substitution}):
\begin{equation}
\label{Schwarz}
\left\{\xi(z),z\right\}=
\frac38\left[\frac{1}{z^2} + \frac{1}{(z-x)^2} + \frac{1}{(z-1)^2}\right]
-
\frac14\left[\frac{1}{z(z-x)} + \frac{1}{z(z-1)} + \frac{1}{(z-x)(z-1)}\right].
\end{equation}
Using (\ref{elliptic:substitution}) and (\ref{second:derivative}) we can rewrite equation (\ref{class:1})
in the form of a Schr\"odinger equation
\begin{eqnarray}
\label{class:2}
\left[-\frac{d^2}{d\xi^2}
-
\tilde U(\xi)
\right]
\tilde\chi(\xi)
=
4x(x-1){\cal C}(x)\tilde\chi(\xi)
\end{eqnarray}
with the (double periodic in $\xi$) potential
\begin{eqnarray}
\nonumber
\label{tilde:U}
 \tilde U(\xi)
& = &
\left.
\Big(\xi'(z)\Big)^{-2}
\left[
U(z) - \frac12\{\xi(z),z\}
\right]
\right|_{z = z(\xi)}
\\[6pt]
& = &
\left(\frac14- \lambda_1^2\right)\left(\frac{x}{z(\xi)} -1\right)
+
\left(\frac14- \lambda_2^2\right)\left[\frac{x(x-1)}{z(\xi)-x} + 2x -1\right]
\\
\nonumber
& + &
\left(\frac14- \lambda_3^2\right)\left(1-\frac{1-x}{1-z(\xi)}\right)
+
\left(\frac14- \lambda_4^2\right)(z(\xi)-x)
+x - \frac12.
\end{eqnarray}
Continuing analytically the function $\xi(z)$ along the closed path encircling the points $0$ and $x$ one gets:
\begin{eqnarray*}
\xi\left({\rm e}^{2\pi i}z\right)
& = &
\xi(z) +\int\limits_0^1 \frac{dt}{\sqrt{t(1-t)(1-xt)}}
= \xi(z) + 2K(x)
\end{eqnarray*}
where $K(x)$ is the complete elliptic integral of the first kind:
\begin{eqnarray*}
 K(x) \equiv \int\limits_0^1 \frac{dt}{\sqrt{(1-t^2)(1-xt^2)}}.
\end{eqnarray*}
The monodromy condition (\ref{monodromy:condition}) thus takes the form
\begin{equation}
\label{monodromy:2}
\tilde\chi^{\pm}\Big(\xi+2K(x)\Big) = {\rm e}^{\pm i \pi \lambda} \tilde\chi^{\pm}(\xi).
\end{equation}

We shall solve (\ref{class:2}) in the large $\lambda $ limit using a standard perturbative method.
First,  assume that
\begin{equation}
\label{zeroord:assumpt}
\tilde U(\xi)  = o\big({\cal C}(x)\big)
\end{equation}
so that in the leading order we can neglect in (\ref{class:2}) the potential term and
the solutions are just plane waves:
\begin{equation}
\label{zeroth:order}
\tilde\chi^{\pm}_0(\xi) = {\rm e}^{\pm i p \xi},
\qquad
p^2 = 4x(x-1){\cal C}(x).
\end{equation}
On the other hand  the monodromy condition (\ref{monodromy:2}) implies
\begin{eqnarray*}
 {\rm e}^{\pm 2i p K(x)} = {\rm e}^{\pm i \pi \lambda}
\qquad \Rightarrow \qquad
p = - \frac{\pi\lambda}{2K(x)},
\end{eqnarray*}
what also proves the consistency of (\ref{zeroord:assumpt}). Hence, in the leading order one obtains:
\begin{equation}
\label{C:zero}
{\cal C}^{(0)}(x) = \frac{\pi^2\lambda^2}{16 x(x-1)K^2(x)}.
\end{equation}
The first correction ${\cal C}^{(1)}(x)$ is given by:
\begin{eqnarray}
\label{first:order:2}
{\cal C}^{(1)}(x)&=&
\frac{-1}{8x(x-1)K(x)}\int\limits_{\xi_0}^{\xi_0+2K(x)}\!\!d\xi\  \tilde\chi_0^{-}(\xi)\tilde U(\xi)\tilde\chi_0^+(\xi)
\\
\nonumber
&=&
\frac{-1}{16x(x-1)K(x)}\int_{[0,x]} \frac{ \tilde U\big(\xi(z)\big) dz}{\sqrt{z(1-z)(x-z)}},
\end{eqnarray}
where in the first line $\Im\,\xi_0 > 0$ while in the second line $[0,x]$ denotes a positively oriented, closed
contour in the complex $z$ plane, surrounding the points $0$ and $x.$

Integrating one gets:
 \begin{eqnarray*}
&& \hskip -1cm \int_{[0,x]} \frac{\tilde U\big(\xi(z)\big)dz}{\sqrt{z(1-z)(x-z)}}
=  \Bigg\lbrace
\left( 1-4  \lambda^2_1\right)  (I_1 - K(x)) + \left( 1-4  \lambda^2_2\right)
 \left( I_2 + (2x -1) K(x)\right)
\\ [4pt]
&& +\left( 1-4  \lambda^2_3 \right)  ( K(x) - I_3)  + \left( 1-4  \lambda^2_4\right)  (I_4 - x K(x))
+ 4 \left( x - \frac12\right)  K(x)
\Bigg\rbrace,
\end{eqnarray*}
where
\begin{eqnarray*}
 I_1 &=& \frac14 \int_{[0,x]} \frac{x dz}{z \sqrt{z(1-z)(x-z)}} =  K(x) - E(x), \\ [4pt]
 I_2 &=&  \frac14 \int_{[0,x]} \frac{x (1-x) dz}{(z-x) \sqrt{z(1-z)(x-z)}} =(1-x) K(x) - E(x), \\ [4pt]
 I_3 &=& \frac14 \int_{[0,x]} \frac{(1-x) dz}{(1-z) \sqrt{z(1-z)(x-z)}} = E(x),\\ [4pt]
I_4 &=& \frac14 \int_{[0,x]} \frac{z dz}{\sqrt{z(1-z)(x-z)}} =   K(x) - E(x).
\end{eqnarray*}
Here $E(x)$ is the complete elliptic integral of the second kind:
\begin{eqnarray*}
E(x)
\; = \;
\int_0^1 \frac{(1-xt^2)\, dt}{\sqrt{(1-t^2)(1-xt^2)}}.
\end{eqnarray*}
The correction to the accessory parameter takes the form:
\begin{eqnarray*}
{ \cal C}^{(1)}(x) =  \frac{-1}{4 x (x-1)}
\left\lbrace
\frac{E(x)}{K(x)}\left( -1 + \lambda^2_1 + \lambda^2_2 + \lambda^2_3 + \lambda^2_4 \right)
+ x \left( 1-  \lambda^2_2 + \lambda^2_4\right ) - \left( \lambda^2_3 + \lambda^2_4\right)
\right\rbrace
\end{eqnarray*}
Since $ {\cal C}(x) = \partial_x f_{\delta\!}\left[^{\delta_3\:\delta_2}_{\delta_4\:\delta_1}\right]\!(x)$ one can calculate the classical block:
\begin{eqnarray*}
 f_{\delta\!}\left[^{\delta_3\:\delta_2}_{\delta_4\:\delta_1}\right]\!(x)
&=& \int  \frac{dx}{4 x (x-1)}
\left\lbrace
\frac{\left( \pi \lambda\right)^2}{4 K^2(x)} \right.
\\
&&
\hskip -10mm
+ \;
\left.
\frac{E(x)}{K(x)}\left( 1 - \lambda^2_1 - \lambda^2_2 - \lambda^2_3 - \lambda^2_4\right)
- x \left( 1-  \lambda^2_2 + \lambda^2_4\right)  + \lambda^2_3 + \lambda^2_4
\right\rbrace
+
{\cal O} \left({1\over \lambda^2}\right).
\end{eqnarray*}
Using
\begin{eqnarray*}
 \int  \frac{dx}{ x (x-1)} \frac{1}{4 K^2(x)}
&=& \frac{1}{\pi} \frac{K(1-x)}{K(x)} \equiv \frac{\tau}{i \pi},  \\ [6pt]
 \int  \frac{dx}{ x (x-1)} \frac{E(x)}{ K(x)} &=& - \frac12 \ln{K^4(x)} - \ln{x},
\end{eqnarray*}
one gets:
\begin{eqnarray*}
f_{\delta\!}\left[^{\delta_3\:\delta_2}_{\delta_4\:\delta_1}\right]\!(x) &=& \frac14 \, \Bigg\lbrace -i \pi \tau \lambda^2
- \frac12 \left( 1 - \lambda^2_1 - \lambda^2_2 - \lambda^2_3 - \lambda^2_4\right) \, \ln{K^4(x)} \\
 [6pt] &&  \ \
- (1 - \lambda^2_2 - \lambda^2_3) \, \ln(1-x) -  (1 - \lambda^2_1 - \lambda^2_2) \, \ln(x)
\Bigg\rbrace  + {\cal O} \left({1\over \lambda^2}\right)
\end{eqnarray*}
or, in terms of $
 \delta =  \frac{1-\lambda^2}{4},
\delta_i =  \frac{1-\lambda^2_i}{4},
$
\begin{eqnarray}
\label{clas_block}
 f_{\delta\!}\left[^{\delta_3\:\delta_2}_{\delta_4\:\delta_1}\right]\!(x) &=& i \pi \tau \left( \delta - \frac{1}{4} \right)
+ \frac12 \left( \frac{3 }{4} - \delta_1 - \delta_2 - \delta_3 - \delta_4\right) \, \ln{K^4(x)}
\\  [6pt]
\nonumber
 &  +& \left( \frac{1}{4} - \delta_2 - \delta_3\right)  \, \ln(1-x)
+  \left( \frac{1}{4} - \delta_1 - \ \delta_2\right)  \, \ln(x)
 + {\cal O} \left({1\over \delta}\right) .
\end{eqnarray}
The absence in the last two formulae of the $x-$independent integration constants follows from the
normalization condition of the block
\(
{\cal F}_{c,\Delta\!}\left[^{\Delta_3\:\Delta_2}_{\Delta_4\:\Delta_1}\right]\!(x).
\)
\section*{Appendix B}
\setcounter{equation}{0}
\renewcommand{\theequation}{B.\arabic{equation}}

Consider $c =\frac32 $ theory with external weights $\Delta_0 = \frac18$.
 For $r \ne s$ all  coefficients $A_{rs}^c\hspace{-3pt}\left[^{\underline{\hspace{3pt}}\,\Delta_0
\;\underline{\hspace{3pt}}\,\Delta_0}_{\hspace{3pt}\,\Delta_0 \;\hspace{3pt}\, \Delta_0} \right] $ are zero.
There are however some non zero terms if $r=s$:
\begin{eqnarray} \label{Arr}
(16)^{\frac{r^2}{2}}\ \mathcal{A}_{rr}^{\hat c=1}\left[^{*\Delta_0\: *\Delta_0}_{\hspace{4pt} \Delta_0\: \hspace{4pt}\Delta_0}\right]
=  \, \left\{
\begin{array}{lcl}
- r^2 \qquad & \mathrm{if} & \quad r \in 2\mathbb{N},
\\[4pt]
2 \qquad & \mathrm{if} & \quad r \in 2\mathbb{N}+1.
\end{array}
\right.
\end{eqnarray}
Moreover, $\Delta_{rr}=0$.
One can show that all elliptic blocks (\ref{H0blocks1}), (\ref{H0blocks2}) satisfy recursion relations (\ref{Hrek}) with corresponding
$g^{1,\frac12}$ functions (\ref{gfunctions}).
Indeed, for the blocks:
\begin{eqnarray*}
&& \mathcal{H}^{1}_{\Delta}\! \left[^{\Delta_0 \; \Delta_0}_{\Delta_0 \; \Delta_0} \right]\!(x)
= g^{1}(x),
\qquad \ \
\mathcal{H}^{1}_{\Delta}\! \left[^{\Delta_0 \; *\Delta_0}_{\Delta_0 \;\hspace{4pt}\Delta_0}  \right]\!(x)
= g^{1}_{*}(x),
\\[6pt]
&&
\mathcal{H}^{\frac12}_{\Delta}\! \left[^{\Delta_0 \;*\Delta_0}_{\Delta_0 \;\hspace{4pt}\Delta_0} \right]\!(x)
=  g^{\frac12}_{*}(x),
\qquad
\mathcal{H}^{\frac12}_{\Delta}\! \left[^{*\Delta_0 \;*\Delta_0}_{\hspace{4pt} \Delta_0 \;\hspace{4pt}\Delta_0} \right] \!(x)
=  g^{\frac12}_{**}(x),
\end{eqnarray*}
the relation (\ref{Hrek}) holds because  all the residues at $\Delta = \Delta_{rs}$ are zero. In the other
cases the formula (\ref{Arr}) becomes helpful:
\begin{eqnarray*}
\mathcal{H}^{\frac12}_{\Delta}\!
\left[^{\Delta_0\:\Delta_0}_{\Delta_0\:\Delta_0}\right]\!(x) &=&
\sum_{r\in 2{\mathbb N}}
(16q)^{\frac{r^2}{2}}
\frac{A_{rr}^c\hspace{-3pt}\left[^{*\Delta_0 \;*\Delta_0}_{\hspace{3pt}\,\Delta_0 \;\hspace{3pt}\, \Delta_0} \right]}{\Delta }
\mathcal{H}^{\frac{1}{2}}_{\frac{r^2}{2}}\!
\left[^{\Delta_0\:\Delta_0}_{\Delta_0\:\Delta_0}\right]\!(x) \\
[4pt]
&+& \sum_{r\in 2{\mathbb N}+1} (16q)^{\frac{r^2}{2}}
\frac{A_{rr}^c\hspace{-3pt}\left[^{*\Delta_0 \; *\Delta_0}_{\hspace{3pt}\,\Delta_0
\;\hspace{3pt}\, \Delta_0} \right]}{\Delta }
\mathcal{H}^{1}_{\frac{r^2}{2}}\!
\left[^{\Delta_0\:\Delta_0}_{\Delta_0\:\Delta_0}\right]\!(x) \\[6pt]
&=& \frac{1}{\Delta} \sum_{r\in 2{\mathbb N}} q^{\frac{r^2}{2}} \ (-r^2) \ \mathcal{H}^{\frac12}_{\frac{r^2}{2}}\!
\left[^{\Delta_0\:\Delta_0}_{\Delta_0\:\Delta_0}\right]\!(x)
+ \frac{2}{\Delta} \sum_{r\in 2{\mathbb N}+1} q^{\frac{r^2}{2}}\  \theta_3(q^2).
\end{eqnarray*}
Substituting    $\mathcal{H}^{\frac12}_{\frac{r^2}{2}}\!
\left[^{\Delta_0\:\Delta_0}_{\Delta_0\:\Delta_0}\right]\!(x) = \frac{2}{r^2}\ \theta_2(q^2)  $
and using the definitions of the theta functions:
\begin{eqnarray*}
\theta_2(q^2) &=& \sum_{n=- \infty}^{\infty} q^{\frac{(n+1)^2}{2}} = 2 \sum_{n=0}^{\infty} q^{\frac{(n+1)^2}{2}},
\qquad
\theta_3(q^2) = \sum_{n=- \infty}^{\infty} q^{2n^2} = 1+  2 \sum_{n=1}^{\infty} q^{2n^2},
\end{eqnarray*}
one gets:
\begin{eqnarray*}
\mathcal{H}^{\frac12}_{\Delta}\!
\left[^{\Delta_0\:\Delta_0}_{\Delta_0\:\Delta_0}\right]\!(x)
&=&
- \frac{2}{\Delta}\sum_{r\in 2{\mathbb N}} q^{\frac{r^2}{2}} \theta_2(q^2)
+ \frac{2}{\Delta} \sum_{r\in 2{\mathbb N}+1} q^{\frac{r^2}{2}}  \theta_3(q^2) \\
& = &
-
\frac{1}{\Delta}\ (\theta_3(q^2) - 1)\ \theta_2(q^2) + \frac{1}{\Delta} \theta_2(q^2)\ \theta_3(q^2) = \frac{1}{\Delta}\ \theta_2(q^2).
\end{eqnarray*}
The last block in (\ref{H0blocks2}), $\mathcal{H}^{1}_{\Delta}\!
\left[^{*\Delta_0\: *\Delta_0}_{\hspace{4pt} \Delta_0\: \hspace{4pt} \Delta_0}\right]\!(x),$
also satisfies the recursion relation:
\begin{eqnarray}
 \label{H**}
\nonumber
&& \hskip -1cm \mathcal{H}^{1}_{\Delta}\!
\left[^{*\Delta_0\: *\Delta_0}_{\hspace{4pt} \Delta_0\: \hspace{4pt} \Delta_0}\right]\!(x)
=
\theta_3(q^2)
+ \sum_{r\in 2{\mathbb N}}  \left( \frac{-r^2}{\Delta} \right)  q^{\frac{r^2}{2}} \,
\mathcal{H}^{1}_{\frac{r^2}{2}}\!
\left[^{*\Delta_0\: *\Delta_0}_{\hspace{4pt} \Delta_0\: \hspace{4pt} \Delta_0}\right]\!(x)
\\ [6pt]
\nonumber
&& + \sum_{r\in 2{\mathbb N}+1} S_{rr}(*\Delta_0) \left( \frac{2}{\Delta}\right)  q^{\frac{r^2}{2}} \,
\mathcal{H}^{\frac12}_{\frac{r^2}{2}}\!
\left[^{*\Delta_0\: *\Delta_0}_{\hspace{4pt} \Delta_0\: \hspace{4pt} \Delta_0}\right]\!(x)
\\ [6pt]
\nonumber
&& = \,
\theta_3(q^2)
- \frac{2}{\Delta} \left(
\sum_{r\in 2{\mathbb N}} q^{\frac{r^2}{2}}\, \theta_3(q^2)
- \sum_{r\in 2{\mathbb N}+1} q^{\frac{r^2}{2}} \, \theta_2(q^2)
\right)
 \left(
 - q\, \theta_3^{-1} \frac{\partial}{\partial q} \theta_3(q) + \frac{\theta_2^{4}(q)}{4 }
 \right)
\\
\nonumber
&& -  \frac{2}{\Delta} \sum_{r\in 2{\mathbb N}} \frac{r^2}{2}\, q^{\frac{r^2}{2}}\, \theta_3(q^2)
+ \frac{2}{\Delta} \sum_{r\in 2{\mathbb N}+1} \frac{r^2}{2}\, q^{\frac{r^2}{2}} \, \theta_2(q^2)
\\ [6pt]
&& = \,
\theta_3(q^2)\left(1  -  \frac{q}{\Delta} \theta_3^{-1}(q) \frac{\partial\theta_3(q)}{\partial q}  + \frac{\theta_2^{4}(q)}{4 }
 \right)
\\ [6pt]
\nonumber
&& + \frac{1}{\Delta}   \left(
 q \theta_3^{-1}(q) \frac{\partial\theta_3(q)}{\partial q}  - \frac{\theta_2^{4}(q)}{4 }
 - \frac{q}{2 }  \frac{\partial}{\partial q}  \right)
\left(
\theta_3^{2}(q^2) - \theta_2^{2}(q^2)
\right).
\end{eqnarray}
From the identities
\begin{eqnarray*}
&& \theta_3(q^2) = \theta_3(q) \left( \frac{1 + \sqrt{1-x}}{2} \right)^{\frac12} , \qquad
\theta_2(q^2) = \theta_3(q) \left(\frac{1 - \sqrt{1-x}}{2} \right)^{\frac12},
\end{eqnarray*}
it follows that
\begin{eqnarray*}
\theta_3^{2}(q^2) - \theta_2^{2}(q^2) =  \sqrt{1-x}\, \theta^2_3(q).
\end{eqnarray*}
Since
\[
\frac{dq(x)}{dx} = \frac{\pi^2 q(x)}{4x(1-x) K^2(x)} = \frac{q(x)}{x(1-x)\theta_3^4(q)}
\]
we  have
\begin{eqnarray*}
q\, \frac{\partial}{\partial q} = x (1-x)\, \theta_3^{4}(q) \frac{\partial}{\partial x}
\end{eqnarray*}
and with the help of the relation
\(
x = \frac{\theta_2^{4}(q) }{\theta_3^{4}(q) }
\)
we finally get
\begin{eqnarray*}
\frac{q}{2}\, \frac{\partial}{\partial q}
\left(
 \sqrt{1-x}\, \theta^2_3(q)
\right)
 = \left(
 q\theta_3^{-1}(q)\frac{\partial\theta_3(q)}{\partial q}
- \frac{\theta_2^{4}(q)}{4 }
\right) \,
\sqrt{1-x} \, \theta^2_3(q),
 \end{eqnarray*}
what demonstrates that the last line in (\ref{H**}) indeed vanishes.

\end{document}